# Sub-Doppler spectroscopy of quantum systems through nanophotonic spectral translation of electro-optic light

**Authors:** David A. Long,[1,*] Jordan R. Stone,[1,2] Yi Sun[1], Daron Westly[1], and Kartik Srinivasan[1,2,*]

**Affiliations:**
  [1]National Institute of Standards and Technology, Gaithersburg, MD 20899, USA
  [2]Joint Quantum Institute, NIST/University of Maryland, College Park, MD 20742, USA
  *Corresponding authors: D. A. Long david.long@nist.gov, K. Srinivasan kartik.srinivasan@nist.gov

**Abstract:**

An outstanding challenge for deployable quantum technologies is the availability of high-resolution laser spectroscopy at the specific wavelengths of ultranarrow transitions in atomic and solid-state quantum systems. Here, we demonstrate a powerful spectroscopic tool that synergistically combines high resolution with flexible wavelength access, by showing that nonlinear nanophotonics can be readily pumped with electro-optic frequency combs to enable highly coherent spectral translation with essentially no efficiency loss. Third-order ($\chi^{(3)}$) optical parametric oscillation in a silicon nitride microring enables nearly a million optical frequency comb pump teeth to be translated onto signal and idler beams; while the comb tooth spacing and bandwidth are adjustable through electro-optic control, the signal and idler carrier frequencies are widely tuneable through dispersion engineering. We then demonstrate the application of these devices to quantum systems, by performing sub-Doppler spectroscopy of the hyperfine transitions of a Cs atomic vapor with our electro-optically-driven Kerr nonlinear light source. The generality, robustness, and agility of this approach as well as its compatibility with photonic integration are expected to lead to its widespread applications in areas such as quantum sensing, telecommunications, and atomic clocks.

The on-chip generation of coherent light across the visible and near-infrared wavelength regions has been a long-standing challenge in integrated photonics[1] and would offer wide-spread use in areas such as atom-based sensors[2], quantum communications[3], and precision measurements[4]. While there is ongoing work on development of heterogeneously integrated lasers[5] and hybrid integrated lasers[6,7] for accessing these wavelength ranges[8], nonlinear nanophotonics provides another route via wavelength conversion. Nonlinear microresonators, in particular, have shown tremendous potential for chip-scale coherent light generation through optical parametric oscillation (OPO)[9-15]. The use of the third-order ($\chi^{(3)}$) nonlinearity allows for much of the visible and near-infrared regions to be accessed using a single 780 nm pump[9,11,12], with the microresonator geometry dictating phase- and frequency-matching and hence the specific output colours generated. This allows optical parametric oscillation (OPO) devices to access the atomic and molecular transitions which are critical for many applications in chemical detection[16], atomic clocks[17], and quantum science[18], all while utilizing a single pump laser technology. However, while OPO may be tuned both thermally and by varying the optical pump wavelength, reliably reaching narrow atomic transitions and performing sub-Doppler spectroscopy – a prerequisite for the use of this technology in many applications – has proven challenging.

Here, we demonstrate the ability to seamlessly combine the high-throughput and high spectral resolution provided by electro-optic combs[19-22] with the wavelength access provided by Kerr nonlinear optics[23]. Utilizing an electro-optic frequency comb as the OPO pump laser enables robust and efficient generation of signal and idler optical frequency combs (see Fig. 1a-b); where the comb tooth spacing and bandwidth are adjustable through electro-optic control, while the signal and idler carrier frequencies are widely tuneable through dispersion engineering. This allows for the on-chip conversion of optical frequency combs over a wide wavelength range in the visible and near-infrared, including at wavelengths where electro-optic modulators may not be readily available. In addition, the resulting optical frequency combs enable spectroscopy on narrow atomic transitions in a multiplexed fashion that alleviates the wavelength tuning requirements on the OPO that would be needed to resolve sub-Doppler features with high temporal resolution. Finally, while other nonlinear processes (e.g., optical harmonic generation[24], sum- and difference-frequency generation[25], or stimulated four-wave mixing[26]) could also be used to spectrally translate electro-optic frequency combs, the use of $\chi^{(3)}$ OPO offers greater wavelength flexibility and only requires a single input laser.

The $Si_3N_4$ microrings utilized here rely on $\chi^{(3)}$ nonlinearity to convert pump laser light into signal and idler waves with frequencies $v_s$ and $v_i$, respectively, that obey the energy conservation condition $2v_p = v_s + v_i$, where $v_p$ is the pump frequency and the microring dispersion determines the signal and idler mode pair[9]. Traditionally, the pump laser is a continuous-wave (CW) source with a linewidth much narrower than the cavity resonance; hence, in this case the OPO emits CW signal and idler waves. Still, modelling of the intracavity nonlinear dynamics (see Methods) indicates that phase modulation of the pump laser is coherently mapped to the signal and idler waves if the modulation is sufficiently slow compared to the cavity photon lifetime, even in a strongly driven regime where the residual carrier power is below the threshold for parametric oscillation (see Fig. 1c). Functionally, such mapping underlies the coherent spectral translation of electro-optic frequency combs – which can be generated through pure phase modulation of a CW laser – that we report here. The fact that the pump comb can drive OPO at a carrier power below the parametric oscillation threshold suggests that the underlying nonlinear physics is more complex than spectral translation of electro-optic comb teeth mediated by the pump/signal/idler tones of a CW OPO. Finally, unlike recently reported results on electro-optic comb conversion in a bulk crystal $\chi^{(2)}$ OPO[27], here we observe that the comb is transferred to both the signal and idler outputs.

To produce the electro-optic frequency comb pump laser, light from an external cavity-diode laser at 780 nm was passed through an electro-optic phase modulator (EOM) which was driven by a frequency chirped waveform (Fig. 1b). The use of a frequency chirp (corresponding to a quadratic phase evolution) has been shown to be an optimal approach for the generation of ultra-flat optical frequency combs with much lower drive voltages than overdriving a phase modulator with a single frequency waveform[19,28]. The frequency chirp was generated by a low-cost direct digital synthesis chip[20], where the resulting frequency comb had a comb tooth spacing given by the repetition rate of the chirped waveform and a comb bandwidth given by twice the chirp range. The optical frequency comb was then passed through a tapered amplifier which produced 30 mW of on-chip power before being injected into the integrated microring OPO (see Methods) through a silicon nitride ($Si_3N_4$) waveguide. While in this work the electro-optic comb was generated using a conventional commercial EOM, recent developments with electro-optic modulators based on thin-film lithium niobate modulators[29], including at 780 nm, suggest a path for on-chip integration[30]. In particular, different types of electro-optic combs have been demonstrated on-chip[31-33], including recent work on frequency-agile, low voltage electro-optic combs interfaced with silicon nitride photonics[34].

Figure 2a shows optical spectrum analyser traces from three OPO devices pumped with the 780 nm electro-optic comb, where variation of the pumped mode results in a significant change in output wavelengths, highlighting the ability to use Kerr nonlinear optics to access the broad range of wavelengths associated with the different quantum systems depicted in Fig. 1a. The generated OPO signal wavelengths are at 703 nm, 665 nm, and 628 nm, while the corresponding idler wavelengths are at 877 nm, 938 nm, and 1018 nm. The first OPO device yields around 85 µW and 30 µW of output idler and signal light, respectively, into a single-mode optical fibre after optical filtering. The signal and idler power was independent of the optical frequency comb bandwidth and tooth spacing, with no observable change in power seen even when the direct digital synthesizer, which produces the optical frequency comb, was turned off. We focus on this OPO (and others like it) going forward, in exploring the properties of the spectrally translated electro-optic comb teeth and the tailoring and application of these devices to Cs spectroscopy.

The tooth spacings of the spectrally translated frequency combs are too fine to be resolved in the optical spectrum analyser. To directly examine the translated EO comb spectrum and its relationship to that of the pump EO comb, we utilized a titanium-sapphire laser to serve as a local oscillator for down-conversion into the radiofrequency domain. As can be seen in the Extended Data Table S1, only slight variations in the power distribution of the frequency comb between negative-order, positive-order, and carrier were observed as the comb was amplified and then passed through the optical chip. The idler showed some reduction of the power in the positive and negative order sidebands relative to the carrier, as would be expected as the optical frequency comb (which was 2200 MHz wide) is wider than the microring's pump mode linewidth of 700 MHz (full-width at half-maximum). However, as can be seen in Fig. 2b, the envelope of the optical frequency comb is only slightly affected by the OPO wavelength conversion, with a flat comb observed at both 780 nm (the pump) and 877 nm (the idler). Critically, this comb flatness enables the high signal-to-noise sub-Doppler spectroscopy that we show later in the paper. We further expect similar behaviour for the signal since the pump and idler behaviour have been characterized, and the three are fundamentally linked through energy conservation (we could not measure the signal directly in the same way as the idler since its wavelength was outside of the titanium-sapphire range).

To visualize the signal comb, we recorded the heterodyne spectrum when it is spectrally separated from the pump and idler and detected with a fast photodiode (see Fig. 3). While this approach does not allow for separation of the positive and negative order comb teeth (as they occur at identical frequencies in the radiofrequency domain) we can readily assess the flatness and coherence of the optical frequency combs. Similarly flat optical frequency combs are observed for the pump, signal, and idler beams. In addition, as can be seen in Fig. 3, the optical frequency comb spacing of the pump can be agilely varied over many orders of magnitude from tens of megahertz (and above) down to 1 kHz (and below), with combs of the same repetition rate then observed at the signal and idler wavelengths. We note that the shown 1 kHz comb had nearly one million individual comb teeth, and the ability to create such a fine-toothed comb is critical for ultra-high resolution spectroscopy[20].

Performing optical frequency comb spectroscopy with a single pump laser requires a local oscillator to be simultaneously injected into the OPO (see Fig. 4a). In this case we split the OPO pump laser's output into probe and local oscillator paths in a self-heterodyne configuration[19]. A comb was generated on the probe leg as previously described above. The local oscillator path was frequency shifted through the use of an acousto-optic modulator in order to break the degeneracy between the positive and negative order comb teeth in the radiofrequency domain. These two paths were then combined, amplified, and injected into the OPO. Despite the presence of two strong tones (the comb carrier and the local oscillator) as well as a multitude of comb teeth, the OPO state was still readily reached and available to perform atomic spectroscopy.

To realize OPO suitable for atomic spectroscopy, we apply the optical parametric oscillation using selective splitting in undulated microrings (OPOSSUM) approach[12], where a photonic bandgap is introduced to engineer dispersion, to our microring design (see Methods). OPOSSUM allows us to target specific wavelengths (e.g., corresponding to atomic transitions) with the OPO, and it supresses competitive Kerr nonlinear processes that are observed in conventional microrings. Here, we design the OPO idler wavelength to be 852 nm for the purpose of probing the $^{133}$Cs $D_2$ transition. High-resolution spectroscopy of the $D_2$ transition in Cs vapor (and more generally, the $D_1$ and $D_2$ transitions in alkali vapors) has been the basis of many quantum technologies, including atomic clocks[35], slow light[36], and high-speed optical quantum memories[37], and thus serves as a useful test case for our hybrid electro-optic/Kerr nonlinear optical technology. Importantly, OPOSSUM enables wide frequency tuning of the OPO (actuated by the pump laser frequency), which can compensate for small discrepancies between the targeted and realized idler wavelength. In our experiments, we readily tune the idler wavelength into alignment with caesium (approximately 852.36 nm in vacuum), and we observe a tuning range >100 pm on either side of this target, as shown in Fig. 4b.

In Fig. 4c we display the energy level diagram for the $^{133}$Cs $D_2$ transition near 852 nm that we probe. Given the relatively high power available (80 µW to 170 µW into the vapor cell), we were able to utilize the strong local oscillator tone provided by the acousto-optically shifted pump to induce saturation in the Cs atoms, thereby allowing the probe electro-optic comb to perform sub-Doppler spectroscopy. As can be seen in Figs. 4d and 4e, the spectrally translated optical frequency comb was then able to rapidly record the Cs Doppler envelope as well as the hyperfine pumping features with a resolution (comb tooth spacing) as high as 100 kHz and a measurement time as short as 1 ms. The high degree of comb flatness gives rise to a high signal-to-noise spectrum which otherwise would be limited by the weakest comb teeth.

The spectral locations of the observed transitions match the expected locations based on the energy level diagram. While in Fig. 4 we probe the hyperfine states associated with transitions from the F=3 ground state, in the Extended Data we show similar spectroscopy of transitions from the F=4 ground state, where the OPO idler output frequency was red-shifted by ≈9 GHz in the manner described above in order to probe this set of states.

Nanophotonic optical parametric oscillation allows for spectral translation from convenient wavelengths where photonic components are low cost and readily available to nearly arbitrary visible and near-infrared wavelengths. Here we have shown that this spectral translation can be performed on an electro-optic frequency comb, providing an approach for reaching and interrogating the narrow resonances of atomic clocks and quantum systems. This approach is highly synergistic, with the electro-optics controlling the behaviour within a given cavity resonance and the Kerr nonlinearity guiding the behaviour across $\approx 10^5$ resonances. While we explicitly demonstrate the application of this hybrid electro-optic/Kerr nonlinear optical technology to sub-Doppler spectroscopy of Cs vapor at 852.3 nm, the basic methodology we have been employed can be straightforwardly extended to any of the wavelengths reached in Fig. 2a, which through fine tuning based on the OPOSSUM approach, can target neutral strontium (698 nm), the nitrogen vacancy centre in diamond (637 nm), and the ground state of epitaxial InAs/GaAs quantum dots (930 nm). Finally, the agility and flexibility of electro-optic frequency combs and Kerr nonlinear optics make these methods likely compatible with other nanophotonic nonlinear optics approaches such as third harmonic generation and Kerr comb generation, with strong applications across much of integrated photonics.

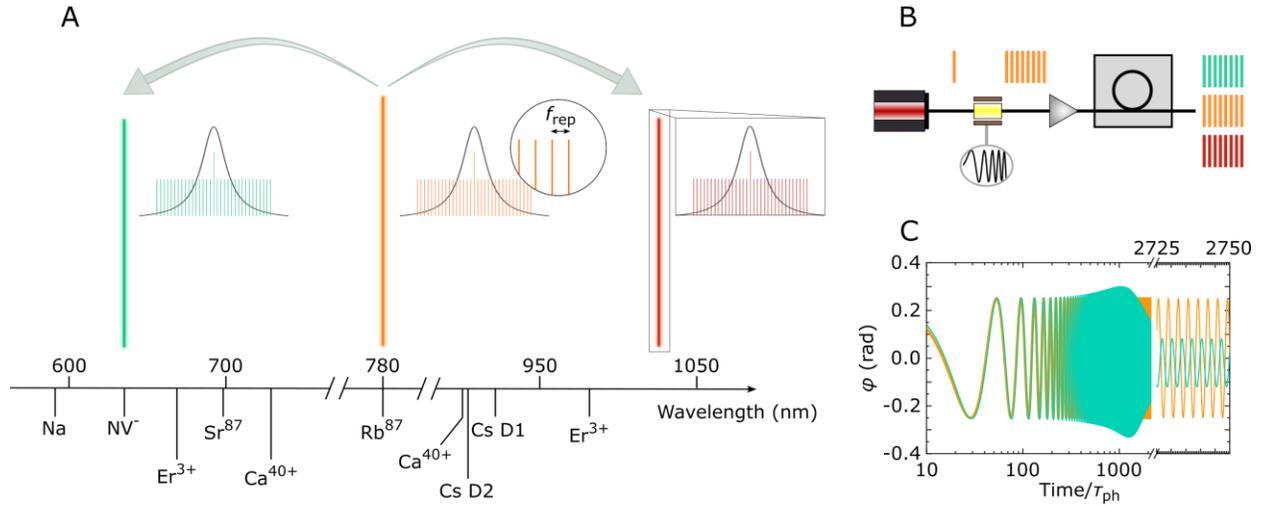

**Figure 1.** Spectral translation of ultrahigh resolution frequency combs via Kerr optical parametric oscillation (OPO). **A.** Depiction of the Kerr OPO spectrum pumped with electro-optic light. Degenerate four-wave mixing (FWM) converts pump laser light (orange) to widely separated signal (green) and idler (red) waves. Hence, the Kerr OPO is useful to generate colors needed for spectroscopy and other applications. The insets accompanying each OPO wave are magnified depictions to illustrate how each wave is an ultrahigh resolution frequency comb with repetition frequency $f_{rep}$, and a bandwidth which exceeds the microring linewidths. **B.** Schematic of our spectral translator. A CW laser is phase modulated using an electro-optic modulator driven by a frequency-chirped microwave source, amplified, and sent to a silicon-nitride microring. **C.** Numerical simulation of coupled mode equations (CMEs) to analyze the phase response of a Kerr OPO signal wave (green) to a phase modulated pump laser (orange). For modulation periods much slower than the photon lifetime, the signal phase preserves any pump phase dynamics, suggesting that faithful transfer of the pump electro-optic comb to the signal should be possible.

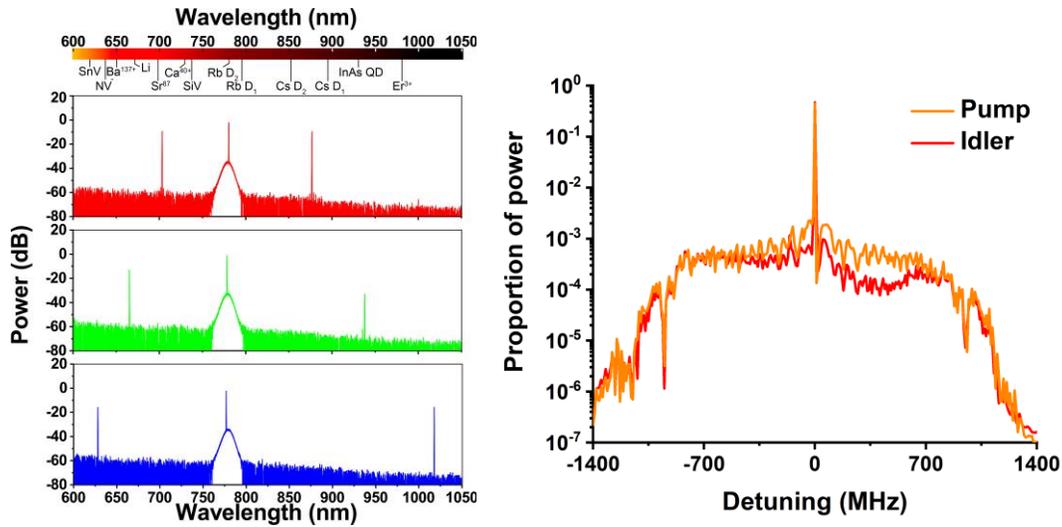

**Figure 2.** Spectral translation of the pump optical frequency comb with the nanophotonic optical parametric oscillator (OPO). (Left panel) Optical spectrum analyser traces showing the pump, signal, and idler outputs of the OPO when driven by an EO comb at the pump frequency (the EO comb bandwidth and tooth spacing are too small to be observed in these spectra). By selecting a different pump mode, the OPO output signal and idler wavelengths can be widely varied. The y-axes are referenced to 1 mW. Amplified spontaneous emission from the 780 nm pump laser system is observed to be 50 dB below the pump power. Also shown are the wavelengths of common quantum systems. (Right panel) Normalized comb tooth powers measured through a beat note with a titanium-sapphire laser for both the 780 nm pump laser (after the photonic chip) and the idler at 877 nm (first spectrum from the left panel). The shown combs had a spacing of 10 MHz and a bandwidth near 2.2 GHz (i.e., 220 comb teeth). The envelope of the idler comb is well matched to that of the incident pump comb.

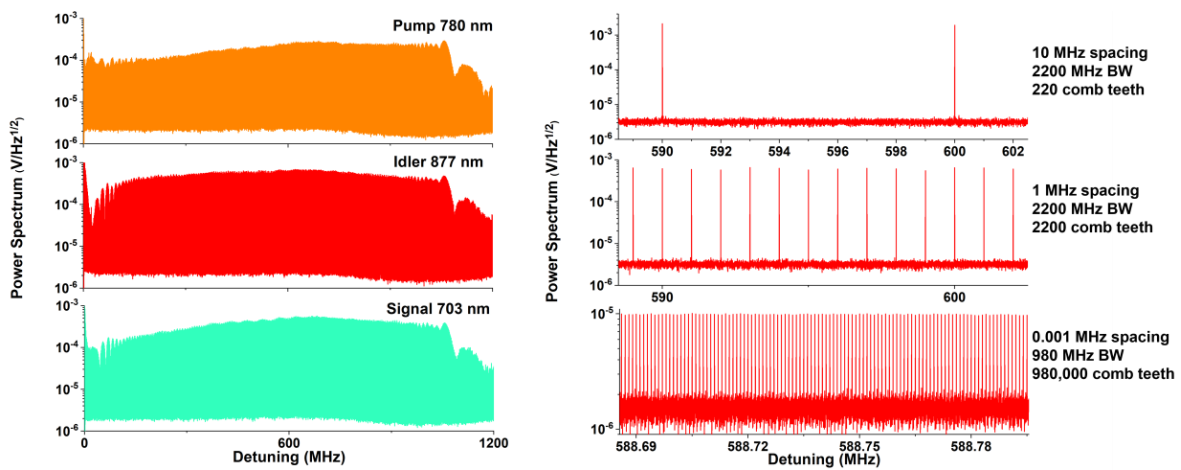

**Figure 3.** Frequency-agile pump, signal, and idler optical frequency combs. (Left panel) Optical frequency comb power spectra for the pump, idler, and signal beams with a comb tooth spacing of 1 MHz. Each of these combs contains roughly 2400 individual comb teeth with a high degree of flatness. (Right panel) Typical idler comb power spectra. The comb spacing can be agilely adjusted over orders of magnitude with narrow comb spacings leading to nearly one million individual comb teeth. In addition, the high coherence of the pump optical frequency combs is readily transferred into the idler and signal combs.

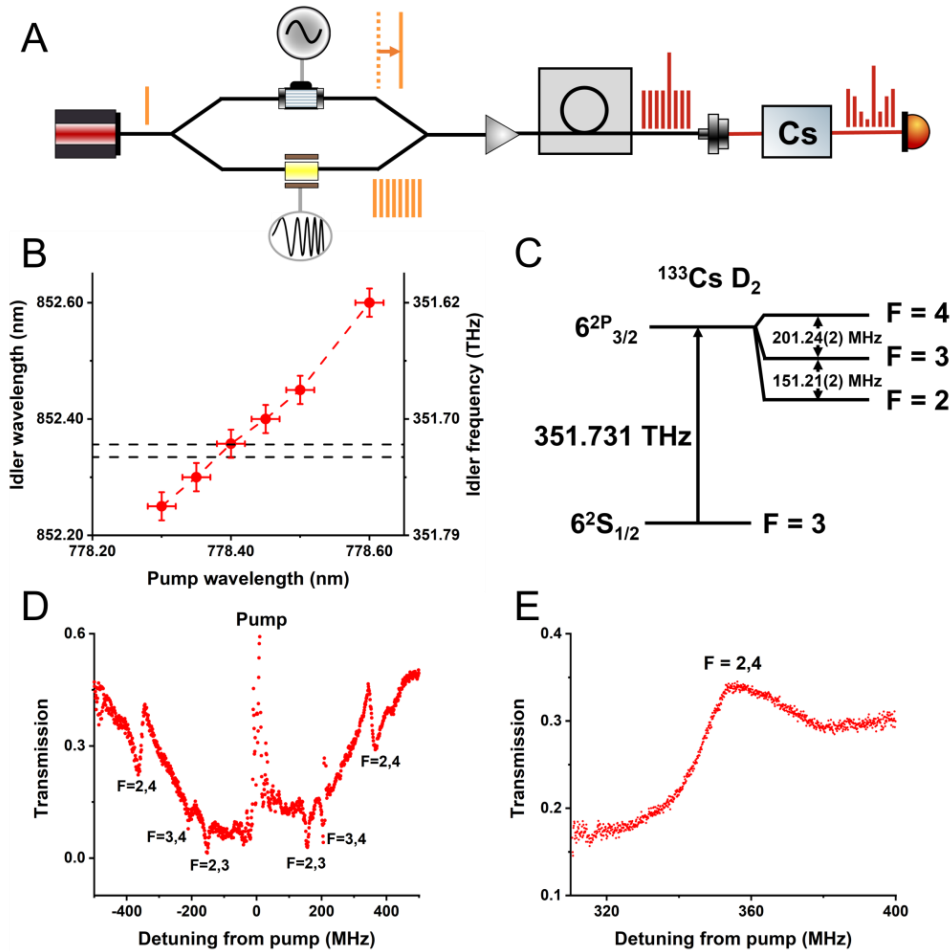

**Figure 4.** Optical parametric oscillator comb spectroscopy of caesium. **A.** Experimental schematic. The laser output was split into two legs. The upper leg was shifted with an acousto-optic modulator to serve as a local oscillator while the lower leg produced an electro-optic frequency comb. These two legs were then combined, amplified, and spectrally translated with the nanophotonic optical parametric oscillator. Following spectral filtration (to remove the signal and depleted pump beams), the idler output was launched into free-space and then sent through a 75 mm room-temperature Cs vapour cell with a beam diameter of ≈6 mm. **B.** Idler wavelength (red data points) versus pump wavelength where the shown uncertainties are the optical spectrum analyser resolution. The dashed black lines mark the two Cs $D_2$ transition frequencies resulting from its ground state splitting. **C.** Relevant energy levels for the hyperfine pumping of a Cs $D_2$ transition[38]. **D.** Optical frequency comb spectrum of the Cs $D_2$ transition which is the average of 10 spectra, each of which was acquired in 1 ms. The shown spectrum includes 1000 individual comb teeth separated by 1 MHz. The local oscillator served as the pump beam and is shown at centre. The six hyperfine pumping transitions are observed at the difference frequencies of the upper state hyperfine levels (see panel C). Their lineshapes have been shown to be strongly dependent upon input polarization[39]. **E.** Cs $D_2$ hyperfine pumping transition recorded with a 100 kHz spaced optical frequency comb containing 900 individual comb teeth. This spectrum is the average of 10 spectra, each of which was acquired in 1 ms.

**Methods:**

**Coupled mode equations:** We model our experimental system by numerically integrating a set of nonlinear coupled mode equations (CMEs):

$$\frac{da}{dt} = \sqrt{\frac{\kappa}{2\hbar\omega_p}} P_{\text{in}} e^{2i\varphi(t)} - \left(\frac{\kappa}{2} + i2\pi\delta\right) a + ig_0(|a|^2 + 2|b_s|^2 + 2|b_i|^2)a + 2ig_0 a^* b_s b_i$$

$$\frac{db_s}{dt} = -\left(\frac{\kappa}{2} + i2\pi\left(\delta + \frac{\Delta\nu}{2}\right)\right) b_s + ig_0(2|a|^2 + |b_s|^2 + 2|b_i|^2)b_s + ig_0 a^2 b_i^*$$

$$\frac{db_i}{dt} = -\left(\frac{\kappa}{2} + i2\pi\left(\delta + \frac{\Delta\nu}{2}\right)\right) b_i + ig_0(2|a|^2 + 2|b_s|^2 + |b_i|^2)b_i + ig_0 a^2 b_s^*$$

where $a$, $b_s$, and $b_i$ represent the complex pump, signal, and idler fields, respectively, $\kappa$ is the loaded resonator loss rate in units of radians, $\omega_p$ is the angular pump frequency in radians, $P_{\text{in}}$ is the on-chip (i.e., in the waveguide) pump power in Watts, $\delta$ is the pump-resonator detuning in Hertz, $\Delta\nu$ is the frequency mismatch in Hertz, and $g_0$ is the single-photon Kerr shift in radians. To model electro-optic modulation, we apply a time-dependent pump laser phase shift, $\varphi(t)$.

**OPO design and fabrication:** For the OPO devices studied in Figs. 2 and 3, we choose microring dimensions (ring width, height, and ring radius) to optimize chromatic dispersion via higher-order dispersion engineering, as discussed in Ref. 9, and we create device layouts using the NIST nanolithography toolbox[40]. We deposit stoichiometric SiN ($Si_3N_4$) of 500 nm thickness by low-pressure chemical vapor deposition on top of a 3 micrometer-thick layer of $SiO_2$ on a 100 mm diameter Si wafer. We fit ellipsometer measurements of the wavelength-dependent SiN refractive index and layer thicknesses to an extended Sellmeier model when choosing device dimensions for optimized dispersion. The device pattern is created in positive-tone resist by electron-beam lithography and then transferred to SiN by reactive ion etching using a $CF_4/CHF_3$ chemistry. After cleaning the devices, we anneal them for four hours at 1100 degrees Celsius in $N_2$. Next, we perform a $SiO_2$ liftoff so that the resonator has an air top-cladding for dispersion purposes while the perimeter of the chip is $SiO_2$-clad for better coupling to lensed optical fibres. The facets of the chip are then polished for lensed-fibre coupling. After polishing, the chip is annealed again.

**OPOSSUM.** Optical parametric oscillation using selective splitting in undulated microrings (OPOSSUM), as used in Fig. 4 and Extended Data Fig. 1, is based on a new paradigm for microresonator dispersion engineering, in which a photonic bandgap (introduced through a photonic-crystal modulation of the microresonator ring width, RW) selects specific modes to participate in four-wave mixing (FWM), while the underlying microresonator dispersion is strongly normal, so that other FWM processes are suppressed via energy and momentum non-conservation. Our dispersion simulations, based on finite element methods, predict increasingly normal dispersion for smaller thickness. Hence, we use 400 nm-thick SiN to ensure globally normal dispersion. We choose the photonic crystal grating period to precisely set the idler mode number ($m = 328$), whose corresponding wavelength we desire to be near 852 nm, and the modulation amplitude ($A_m = 5$ nm) and RW (800 nm) are set to balance the underlying (normal) dispersion to ensure phase- and frequency matching. For more details on OPOSSUM design, we refer the interested reader to Ref. [12].

**Data and materials availability:** All data and supporting materials will be made available at a DOI to be provided through the National Institute of Standards and Technology.

**Methods Only References:**

40. Balram, K. C. *et al.* The nanolithography toolbox. *J. Res. Natl Inst. Stand. Technol.* **121**, 464–475 (2016).

**Funding:** This work was supported by the NIST-on-a-Chip and DARPA LUMOS programs. Portions of this research were performed in the NIST CNST NanoFab.

**Author contributions:**

Conceptualization: DAL, JRS, KS

Investigation: DAL, JRS, YS, DW

Funding acquisition: DAL, JRS, KS

Writing – original draft: DAL, JRS

Writing – review & editing: DAL, JRS, KS

**Competing interests:** A provisional patent has been filed on some of these concepts on which several of the authors (DAL, JRS, KS) are listed as inventors.

**Extended Data:**

**Table S1.** Proportion of the total optical power found in the negative order comb teeth (Neg. Order), carrier tone, and positive order comb teeth (Pos. Order) for the 780 nm pump before the tapered amplifier, after the tapered amplifier, and after the microring chip as well as for the 877 nm idler. These measurements were recorded with a comb having

a 10 MHz spacing and a 2200 MHz bandwidth. The shown uncertainties are one standard deviation from a series of repeated measurements.

|  | Pump Before Amp. | Pump After Amp. | Pump After Chip | Idler |
|---|---|---|---|---|
| **Neg. Order** | 0.112(1) | 0.107(7) | 0.109(5) | 0.077(3) |
| **Carrier** | 0.789(2) | 0.790(5) | 0.805(4) | 0.882(3) |
| **Pos. Order** | 0.099(1) | 0.103(9) | 0.086(7) | 0.041(1) |

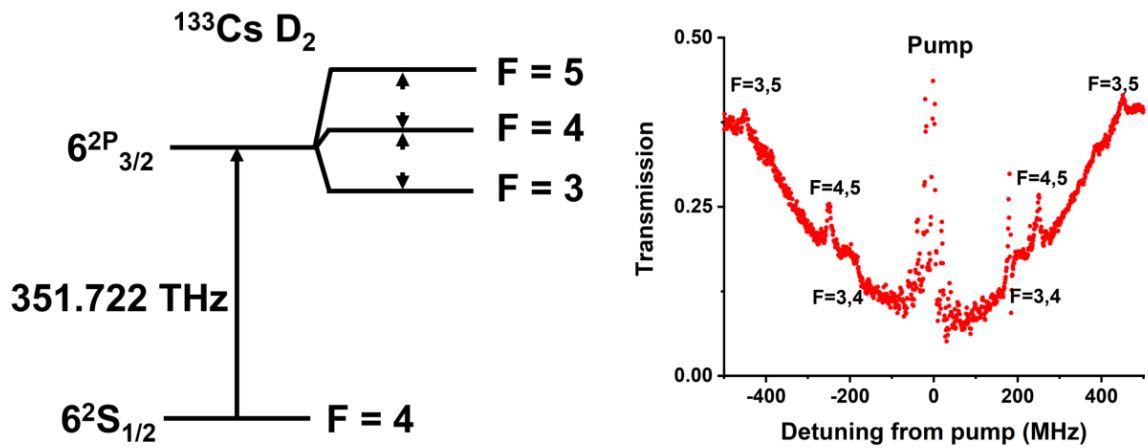

**Figure S1.** Energy level diagram and optical frequency comb spectrum for transitions from the F=4 ground state. (Left panel) Relevant energy levels for the hyperfine pumping of a Cs $D_2$ transition[38] from the F=4 ground state. (Right panel) Optical frequency comb spectrum of the Cs $D_2$ transition which is the average of 10 spectra, each of which was acquired in 1 ms. The six hyperfine pumping transitions are observed at the difference frequencies of the upper state hyperfine levels. The excess noise near 181 MHz (i.e., the AOM frequency for this data set) occurs as those comb teeth are near to DC in the comb FFT.